\documentstyle[preprint,eqsecnum,aps,epsf]{revtex}
\newif\iftightenlines\tightenlinesfalse
\tightenlines\tightenlinestrue
\begin{document}

\title{$Z$ polarization in the Higgs boson signal at the LHC}
\author{M. J. Duncan$^{*}$   and M. H. Reno}
\address{
Department of Physics and Astronomy, University of Iowa, Iowa City,
Iowa 52242}

\maketitle

\begin{abstract}
We examine a method of studying the $Z$ polarization in $H\rightarrow
ZZ\rightarrow \ell^+\ell^-\nu\bar{\nu}$ in proton-proton collisions
at the Large Hadron Collider (LHC). Included are the dominant
contributions: gluon fusion production of the Higgs boson, continuum
production of $Z$ pairs, and the $Z$+missing jet QCD background.
The polarization signal is distinguishable from the background for
Higgs masses less than or equal to $\sim 700$ GeV, for
an integrated luminosity of $10^5$ pb$^{-1}$.
\end{abstract}

\section{INTRODUCTION}

Polarization information from the decays of weak gauge bosons 
provide an important clue to their production mechanism.
In particular, if a weak boson pair is produced from the decay of
a heavy Higgs boson, the bosons are largely longitudinally polarized.
On the other hand, continuum production of weak bosons yields
primarily transversely polarized bosons. Extracting the polarization
signal from weak boson decays will be important to characterize
a Higgs boson signal.
Alternative models with a strongly interacting electroweak symmetry
breaking sector also
have enhanced longitudinal boson pair production rates.\cite{ssb} 
The polarization of the weak bosons may be one way to
differentiate between signal and background.

Polarization effects in Higgs boson decay into the `gold-plated mode'
$H\rightarrow Z Z\rightarrow
\ell_1^+\ell_1^-\ell_2^+\ell_2^-$ 
have been investigated in detail in the literature.\cite{goldplate} 
At the proposed Large Hadron Collider (LHC) with proton-proton collisions at
$\sqrt{S}=14$ TeV and 
an annual integrated luminosity of $10^5$ pb$^{-1}$, polarization information
for high mass Higgs bosons
will be difficult to extract with the limited statistics.
To aid in a Higgs boson search at the LHC, and as a complementary
analysis,
we have proposed 
using the process
$H\rightarrow ZZ\rightarrow \ell^+\ell^-\nu\bar{\nu}$.\cite{dr,dpf}
This decay mode has a factor of six enhancement over the gold-plated
mode when $\ell=e,\ \mu$.\cite{cc}
Our analysis involves the average value of a quantity obtained from
the charged lepton angular decay distribution, which differs for
leptons coming from longitudinally
polarized $Z$'s, ($Z_L$) and transversely polarized $Z$'s ($Z_T$).
The precise quantity is described in Section II.

By considering the average value rather than the decay distribution itself,
there are two advantages. First, the statistics in an experiment at the 
LHC
will be inadequate to examine angular decay distributions as a function
of transverse mass or missing transverse momentum of the event for
heavy Higgs boson decays. 
The average of the lepton angular decay distribution
nevertheless distinguishes between background and signal plus background.
Second, the average value of this quantity is fairly insensitive to
theoretical uncertainties in the differential cross sections related
to scale choices and higher order corrections.

Our goal in this paper 
is to establish the feasibility of this polarization measurement at
the LHC using the process $H\rightarrow ZZ\rightarrow \ell^+\ell^-\nu
\bar{\nu}$. 
Consequently, we consider only the dominant contributions to the
event rate with a charged lepton pair and large missing transverse
momentum. We examine the polarization signal in heavy Higgs production
and decay to charged leptons pairs and missing energy for
10$^{5}$ pb$^{-1}$ of integrated luminosity. Our results suggest that
with sufficient enhancement of longitudinal boson scattering in some
non-resonant strongly interacting vector boson models, the polarization
signals may be observable at high energies.

In the next section, we describe the heavy Higgs boson signal of interest,
$H\rightarrow ZZ\rightarrow \ell^+\ell^-\nu\bar{\nu}$, and the polarization
variable. 
In Section III, we discuss the various Higgs boson production
mechanisms and background processes. 
In Section IV, we show that with the dominant
contributions to the signal, irreducible  and reducible backgrounds,
a Higgs boson signal in the mass range considered here, from 400-800 GeV,
can be extracted except at the highest Higgs masses. A factor of four
increase in integrated luminosity makes polarization signals at $m_H=800$
GeV feasible.
We also show the
effects of experimental cuts and uncertainties associated with theory
and statistics. In Section V, we summarize our conclusions.

\section{HIGGS SIGNAL IN $Z$ POLARIZATION} 

We begin with a discussion of
$H\rightarrow ZZ$.
The Higgs boson mass range considered here is 400-800 GeV. This range
was chosen because for masses lower than $\sim 400$ GeV, the QCD background
rates for $Z+$missing jet are very high. In the mass range
below 400 GeV, the gold-plated mode has an event rate
large enough for unambiguous Higgs identification.
The upper bound was chosen based on unitarity estimates of the maximum
allowed Higgs mass, first done by Lee, Quigg and Thacker.\cite{unitarity}
Over this range of Higgs masses, the branching fraction into $Z$ pairs is
approximately $33\%$, with the remaining width primarily due to $W$ pair
production. 

Essential to our analysis is the fact that the Higgs
boson decays preferentially into longitudinally polarized weak bosons.
A straightforward calculation of the ratio of the polarized widths
gives
\begin{equation}
{\Gamma (H\rightarrow Z_L Z_L)\over \Gamma (H\rightarrow Z_T Z_T)}=
{m_H^4\, (1-2{m_Z^2/ m_H^2})^2\over 8\, m_Z^4 }=38-708
\end{equation}
for masses $m_H=400-800$ GeV. Already at $m_H=500$ GeV, the ratio is 100.
The angular decay distribution of the charged leptons from the
$Z$ decay  characterize the polarization of the $Z$.
Defining the  angle $\theta$
by the $Z$ momentum axis  and the charged lepton three
momentum in the $Z$ rest frame, the angular decay distributions, with
$z\equiv\cos\theta$, are
\begin{eqnarray}
\phi_L(z)&={3\over 4}(1-z^2)\\
\phi_T(z)&={3\over 8} (1+z^2)
\end{eqnarray}
for longitudinally and transversely polarized $Z$'s respectively.
With these distributions, one sees that $\langle |z|\rangle=3/8$ for
purely longitudinal $Z$'s and $\langle |z|\rangle=9/16$ for 
purely transverse $Z$'s. This is the effect that we exploit.

The large enhancement factor in the 
ratio of longitudinal to transverse polarized
$Z$'s is very promising, but
unfortunately, the longitudinal polarization four-vector
is not boost invariant in general. The ratio in Eq. (2.1) is valid for Higgs
decay only in its rest frame. The ratio decreases as one boosts the Higgs
to larger momenta. For gold-plated modes, one can boost back to the Higgs
rest frame and  use the $Z$ momenta in that frame to define the
polarization axes. This has been studied in 
Refs. \cite{malcolm,jjvdb}. 
With one $Z$ decaying to neutrinos,
it is not possible to
unambiguously determine the Higgs rest frame.
The heavy Higgs masses considered here are large enough that the
typical boosts from the Higgs rest frame to the collider
frame are small, so
one still sees a significant enhancement in
the production of $Z$'s with longitudinal polarization in the hadron
center of mass (collider) frame. We use the momentum axis of the reconstructed
$Z$ in the collider  frame to define the angle $\theta$ for the $\ell^-$
angular distribution.

In our numerical results presented below, we evaluate the average value
of $z^*\equiv |z|=|\cos\theta |$.
By constructing the longitudinal polarization four-vector
$\epsilon_L$ in the collider frame from the reconstructed
$Z$ momentum
$p_Z=p_{\ell}+p_{\bar{\ell}}$, namely,
\begin{equation}
\epsilon_L= {1\over M_Z}(|\vec{p}_Z|,E_z{\vec{p}_Z\over|\vec{p}_Z|})\ ,
\end{equation}
one finds for $p_\ell$, the charged lepton momentum in the collider
frame, that
\begin{equation}
z^*={2\over M_Z}|\epsilon_L\cdot p_\ell |\ .
\end{equation}
We evaluate the average value of $z^*$ as a function of transverse
mass $M_T$ where 
\begin{equation}
M_T^2=[(\vec{p}_T\, ^2+m_Z^2)^{1/2}+(\vec{p\llap/}_T\, ^2+m_Z^2)^{1/2}]^2-
(\vec{p}_T
+\vec{p\llap/}_T)^2
\end{equation}
Here, $\vec{p\llap/}_T$ refers to the momentum carried by the neutrino
pair transverse to the beam axis. For muons, the 
experimental evaluation of $\langle
z^*\rangle$ should be very precise. The experimental errors associated
with missing transverse momentum enter into the determination of
the transverse mass, not the evaluation of $z^*$.

Operationally, our 
procedure here is to evaluate the dominant contributions to the total
cross section for
$pp\rightarrow \ell^+\ell^-+p\llap/_T$ and to determine their
average $z^*$ values as a function of transverse mass, then take
the cross section weighted average:
\begin{equation} 
\langle z^*\rangle ={\sum \langle z^*\rangle_i\sigma_i\over \sum \sigma_i}\ .
\end{equation}
We make 
the following cuts on the transverse momentum and rapidity 
of the charged leptons, as well as a cut requiring that the charged
lepton pair reconstruct to a $Z$:
\begin{eqnarray}
&p_T^\ell >20\ {\rm GeV}\\
&|y^\ell |< y_c^{\ell}\\
&M_Z-\Gamma_Z  <M_{\ell^+\ell^-} <M_Z+\Gamma_Z 
\end{eqnarray}
We consider two choices for
$y_c^\ell$: $y_c^\ell = 2.5$ and $y_c^\ell = 3$. 
Increasing the transverse momentum cut for the charged leptons has the effect 
of reducing the separation between the purely longitudinal 
and purely transverse values for
$\langle z^*\rangle$. 
As a practical matter, the invariant mass cut has no effect on our
calculations presented below because the dominant background
contributions include $Z\rightarrow \ell^+\ell^-$ in the
final state. In our calculations,
we use the narrow width approximation
for the $Z$ decay to leptons. The invariant mass cut does
reduce backgrounds from, $e.g.$, $t\bar{t}\rightarrow b\bar{b}W^+W^-
\rightarrow \ell^+\ell^-\nu\bar{\nu} X$. 

In the next section we describe the various contributions to the 
total signal plus background cross
section. In our numerical results, we include only $\ell=\mu$, however,
we do not factor in efficiencies in our evaluation of event rates.
We use the leading order CTEQ3 parton distribution functions and 
five-flavor $\Lambda
= 132$ MeV
in our evaluation
of the cross sections.\cite{cteq}
In terms of the incoming parton momenta $p_1$ and $p_2$, the factorization
and renormalization scales are set to $\mu^2=(p_1+p_2)^2$.

\section{CROSS SECTIONS}

The dominant Higgs signal production mechanism for $\sigma (pp\rightarrow
H\rightarrow ZZ)$ depends on the top quark mass.
The reports of the discovery of the top quark with a mass of $m_t\simeq 175$
GeV\cite{top} mean that 
the largest contribution to $\sigma (pp\rightarrow H)$ at LHC 
energies for
$m_H=400-800$ GeV comes from gluon fusion
$gg\rightarrow H\rightarrow ZZ$, where the gluons couple
to the Higgs
through a triangle diagram with a top quark internal
loop. In principle, one should include 
non-resonant $gg\rightarrow Z_LZ_L$ contributions, however, at
$\sqrt{S}=14$ TeV, these contributions are small.\cite{fullgg} 
Contributions to Higgs production with top quarks in the initial or final
state are suppressed so as to give no appreciable contribution
to the cross section.\cite{gunion} In what follows, we set
$m_t=175$ GeV.

A second important production mechanism for
$Z$-pairs is through vector boson fusion.
Initially, calculations were 
done using the effective $W$ approximation, giving
$WW\rightarrow H$ and $ZZ\rightarrow H$, using the $W$ and $Z$ distributions
in the proton.\cite{dawson} 
More recently, full calculations of $qq\rightarrow 
VVqq\rightarrow Hqq$ and related processes are done instead.\cite{bg}
Using the Higgs resonance portion of $qq\rightarrow ZZqq$
and related processes with quarks and antiquarks in the
initial state, we find that
the cross section from vector boson fusion is 1/10 to 1/3 of the
gluon fusion cross section for the mass range from 400-800 GeV.
In line with our aim to evaluate the feasibility of the polarization
measurement, we
only include the dominant gluon fusion part of the Higgs cross section
and comment below on the effect of changing the normalization of the
signal part of the contribution to $\ell^+\ell^-+
p\llap/ _T$.

The production of $Z$ pairs from quark fusion is the dominant contribution
to the irreducible background to the $H\rightarrow ZZ$ signal in the
Higgs resonance region. 
We include
here the leading order contribution: $q\bar{q}\rightarrow ZZ$, as in
Ref. \cite{malcolm}. 
The transverse and longitudinal $Z$ boson contributions to the total
$q\bar{q}\rightarrow ZZ$ differential cross section, as a function of
the $Z$-pair invariant mass, are shown in Fig. 1. Also indicated is
the $gg\rightarrow H\rightarrow ZZ$ result for $m_H=600$ GeV.
Longitudinal $Z$-pair production is indicated by the dashed lines. As 
advertised, the $q\bar{q}\rightarrow Z_TZ_T$ contribution significantly
dominates the irreducible background. 
The next to leading order
to $q\bar{q}\rightarrow ZZ$, with crossed diagrams, gives an enhancement
of a factor of $\sim 1.2-1.3$ in the cross section.\cite{oo} 
Our Monte Carlo calculation of $pp\rightarrow ZZ$ relies on the leading
order matrix element for the irreducible background, but
we discuss below the consequences of increasing the normalization of this
background.

The QCD process with production of a
single $Z$ plus missing jet is the largest reducible background.
At leading order in $\alpha_s$, this comes from $q\bar{q}\rightarrow
Z g\rightarrow\ell^+\ell^- g$ and crossed diagrams. A fraction of
the events will have the final state parton which is missed in the detector,
so it contributes to events with $\ell^+\ell^-+p\llap/ _T$. We model this
background by imposing the selection cuts outlined in Sec. II, together
with a requirement that the rapidity of the parton $y^p$ satisfy
\begin{equation}
|y^p|>y^p_c.
\end{equation}
The idea here is that for high rapidity partons, 
the parton ``jet'' is outside 
of the detector coverage,
and thus is ``missing.'' We consider two values of $y_c^p$:
$y_c^p=3$ and $y_c^p=4$. The combination of $y_c^p=4$ with $y_c^\ell=2.5$
is particularly difficult to satisfy for this background process.
The rate for single $Z$ production at large
$p_T$ is very high compared to the rate for
$Z$-pairs at the same transverse momentum, so the rapidity cuts
are essential.

The $Z+$missing jet background is the largest reducible background.
Other processes, not included in this analysis, 
have some of the ingredients of the $\ell^+\ell^-\nu
\bar{\nu}$ signal, but typically have additional activity in the
event. For example, the cross section for
$gb\rightarrow Zb
\rightarrow \ell_1^+\ell_1^-c\ell_2^-\bar{\nu}_2$ is large.
When the $b$ decays outside of the central region, it is included in
the $Z+$ missing jet background. When it decays in the central region,
at high transverse momentum of $\ell_1^++\ell_1^-$, the
electron and quark jet will largely align with the missing momentum.
Cuts that veto these events are required. The requirement of no central
jets, together with the invariant mass cut in Eq. (2.10), also
eliminates the background from
$gg\rightarrow t\bar{t}\rightarrow \ell^+\nu_l\ell^-\bar{\nu}_\ell
b\bar{b}$.

To illustrate the dominant signal, irreducible and reducible backgrounds,
we show ${\rm d}\sigma /{\rm d}M_T$ in Figs. 2 and 3. The solid lines
show the gluon fusion contribution to Higgs production. The dashed
line shows $q\bar{q}\rightarrow ZZ$, and the dotted line indicates
the QCD $Z$+missing jet contribution. The fully correlated decays of
one $Z$ to one family of charged leptons, and the other $Z$ to neutrino
pairs, is included in the $ZZ$ cross sections.
The cuts
applied are those in eqs. (2.8-2.10,3.1). In Fig. 2, we have
set $y_c^p=y_c^\ell = 3$. The reducible background is quite large for
$M_T<700$ GeV, making these cuts less than ideal
except for very high mass Higgs bosons. A better choice
is shown in the next figure. 
Fig. 3 shows the same quantities with
$y_c^p = 4$ and $y_c^\ell=2.5$. 

The transverse mass peaks stand out
well for $m_H=400$ GeV and 600 GeV in Fig. 3. 
The event rates are such that the
discovery mode will be the four-charged-lepton final states. 
Using the cuts in Fig. 3
for the $\mu^+\mu^-+p\llap/ _T$ final states, we find 770 signal events for
$m_H=400$ GeV in the range of $M_T=350-450$ GeV and 240 background events
in the same range, assuming an integrated luminosity of 10$^5$ pb$^{-1}$.
For $m_H=600$ GeV, the same cuts yield 160 events for the Higgs signal
and 77 events for the background for $M_T=500-700$ GeV.
An 800 GeV Higgs boson yields 24 signal events and 19 background events
in a range of $M_T=700-900$ GeV.
Using the cuts in Fig. 2, the signal event rates are slightly higher,
but the background event rates are significantly higher except for the
highest mass range. For $M_T=700-900$ GeV and $m_H=800$ GeV, the
signal remains at 24 events and the background is 25 events for the
Fig. 2 cuts.

\section{RESULTS for $\langle z^*\rangle$}

We present our results in a series of figures where we
show the value of $\langle z^*\rangle$ as a function of transverse
mass.  Here,
$\langle z^*\rangle$ is evaluated via Eq. (2.7). Combining
the reducible and irreducible backgrounds, we write Eq. (2.7) as
\begin{equation}
\langle z^*\rangle
={ \langle z^*\rangle_S\sigma_S+\langle z^*\rangle_B\kappa\sigma_B
\over
\sigma_S+\kappa\sigma_B} ,
\end{equation}
for the signal (S) and background (B) differential cross sections
indicated in Figs. 2 and 3 with $\kappa=1$. 

The values of $\langle z^*\rangle$ are subject to uncertainties which
include the relative normalization of the signal to background, as
well as measurement uncertainties. 
A K-factor $\kappa$ in Eq. (4.1) is 
included to estimate the error associated with QCD corrections by
changing the relative normalization of the signal to background
cross sections. A value of $\kappa=1.5$ is a rough estimate of
the theoretical uncertainty in the background calculation.

We begin with
$y_c^\ell=y_c^p=3$, and 
plot $\langle z^*\rangle$
versus $M_T$ for 400, 600 and 800 GeV Higgs bosons in Figs. 4a, 4b and 4c.
The result of Eq. (4.1) with $\kappa=1$ is shown by the solid line and
$\kappa=1.5$ is given by the dotted line. The values of
$\langle z^*\rangle_S$ are indicated by the dot-dashed line,
and $\langle z^*\rangle_B$ are indicated by the dashed line.
The 400 and 600 GeV
Higgs boson transverse mass peaks 
in Fig. 2
translate to
dips in the value of $\langle z^*\rangle$ as the cross section moves
from background dominated to signal dominated.  The broader dip for the case 
of $m_H=800$ GeV is still evident. The reducible $Z+$missing jet background
nearly obscures the dip for $m_H=400$ GeV when 
$y_c^\ell=y_c^p=3$.

We now turn to the rapidity cuts that reduce the $Z+$missing jet background.
In Figs. 5a, 5b and 5c, we show ${\langle z^*\rangle}$ versus
$M_T$ for $y_c^\ell=2.5$ and $y_c^p=4$. These plots 
have error bars with our estimate of the uncertainty in the measurements
of $\langle z^*\rangle$ due to statistics. The error bars were estimated by
using using events generated by
a Monte Carlo generator with the three dominant signal, reducible
and irreducible background contributions. The Monte Carlo events passing
the cuts are grouped by transverse mass bin. Values of $\langle z^*\rangle$
are determined for many collections of $N$ events for a particular bin,
where $N$ is the theoretically predicted number of events in the
bin based on an integrated luminosity of $10^{5}$ pb$^{-1}$. The error
bar for the bin is the standard deviation of these values of
$\langle z^*\rangle$.
In Fig. 5c, we have slightly
offset the central values of the transverse mass bins to better exhibit
the overlapping error bars.

For $m_H=400-600$ GeV, the polarization signal plus background 
is distinguishable from the polarization with no Higgs boson contribution.
For $m_H=800$ GeV, the distinction between $\langle z^*\rangle$ with
and without the Higgs boson is difficult to make with an integrated
luminosity of $10^{5}$ pb$^{-5}$ because of limitations due to the statistical
error. If the event rate can be increased by a factor of four, then this 
method looks more promising. For $m_H=700$ GeV, the cross section is
large enough to reduce the statistical errors and make a 
distinction between background and signal plus background 
values of $\langle z^*\rangle$ feasible.
Since our event rates are determined using
leading order matrix elements and only dominant contributions, 
the error bars presented here may be conservative. However, the error
bars are indicative of how difficult the full angular distribution
measurements will be for heavy Higgs masses.

\section{Conclusions}

Resonant peak searches in the four charged lepton decay modes of
heavy Higgs decay, $H\rightarrow ZZ\rightarrow \ell^+_1\ell^-_1
\ell^+_2\ell^-_2$, remain the preferred path to Higgs boson discovery,
except at the highest Higgs masses considered here.  Before the integrated
luminosity achieves $10^5$ pb$^{-1}$, measurements of $\langle z^*\rangle$
for $m_H=400-600$ GeV
are a reasonable alternative to the full angular distributions
described in Ref. \cite{jjvdb}. In a scenario with $m_H=800$ GeV at
the LHC with $10^5$ pb$^{-1}$, event rates are low and the peak is not
pronounced. A measurement of $\langle z^*\rangle$ supplements a
measurement of an enhanced cross section for $pp\rightarrow ZZ
\rightarrow \ell^+\ell^-\nu\bar{\nu}$, and characterizes the
production mechanism of the $Z$ pair, however an integrated luminosity
of more than  $10^5$ pb$^{-1}$ is required for a measurable difference
between signal plus background and background alone in the
$\langle z^*\rangle$ distribution according to our dominant production
analysis. 
For masses smaller than 800 GeV, this method can characterize the
production mechanism for $Z$-pairs.
Our results 
suggest that this technique may also be applied to study models\cite{ssb} 
with enhanced, non-resonant or broad-resonant, 
longitudinal vector boson scattering.

\acknowledgements
Work supported in part
by National Science Foundation Grants No. PHY-9307213
and PHY-9507688. MHR thanks F. Paige,
Chung Kao  and H. Baer for useful conversations.

\begin{figure}
\caption{Separate contributions to the $M_{ZZ}$ 
differential cross section for $pp\rightarrow ZZ+X$: $q\bar{q}\rightarrow
ZZ$ (solid line), which is the sum of $q\bar{q}\rightarrow Z_TZ_T,\ 
Z_LZ_T$ (dot-dashed lines) and 
$Z_LZ_L$ (dashed lines) and the resonant $gg\rightarrow H
\rightarrow ZZ$ contribution (solid line) and $gg\rightarrow H\rightarrow
Z_L Z_L$ (dashed lines)
for $m_H=600$ GeV. No cuts were applied to these distributions.}
\end{figure}
\begin{figure}
\caption{Differential 
cross section for $pp\rightarrow ZZ+X$ as a 
function of transverse mass $M_T$, with contributions from:
$q\bar{q}\rightarrow ZZ$ (dashed line)
the QCD $Z$+missing jet background (dotted line)
and resonant gluon fusion (solid line)
for $m_H=400$, 600 and 800 GeV. The sum of the three contributions
is indicated with the heavy solid line. The rapidity cuts are
$y_c^p=y_c^\ell = 3$.}
\end{figure} 
\begin{figure}
\caption{Differential 
cross section for $pp\rightarrow ZZ+X$ as a 
function of transverse mass $M_T$, with contributions from:
$q\bar{q}\rightarrow ZZ$ (dashed line)
the QCD $Z$+missing jet background (dotted line)
and resonant gluon fusion (solid line)
for $m_H=400$, 600 and 800 GeV. The sum of the three contributions
is indicated with the heavy solid line. The rapidity cuts are
$y_c^p=4$ and $y_c^\ell = 2.5$.}
\end{figure} 
\begin{figure}
\caption{The values of $\langle z^*\rangle$ versus $M_T$ for
a) $m_H=400$ GeV, b) $m_H=600$ GeV and c) $m_H=800$ GeV, with
$y_c^\ell=y_c^p=3$. The solid line comes from the evaluation
of Eq. (4.1) with $\kappa = 1$, the dotted line, with $\kappa=1.5$. The
Higgs values alone are shown with the dot-dashed line, and the background
alone, with the dashed line. 
}
\end{figure}
\begin{figure}
\caption{The values of $\langle z^*\rangle$ versus $M_T$,
as in Fig. 4, for
a) $m_H=400$ GeV, b) $m_H=600$ GeV and c) $m_H=800$ GeV, with
$y_c^\ell=2.5$ and $y_c^p=4$.
The error bars represent a statistical error  
calculated assuming an integrated
luminosity of $10^5$ pb$^{-1}$.}
\end{figure}

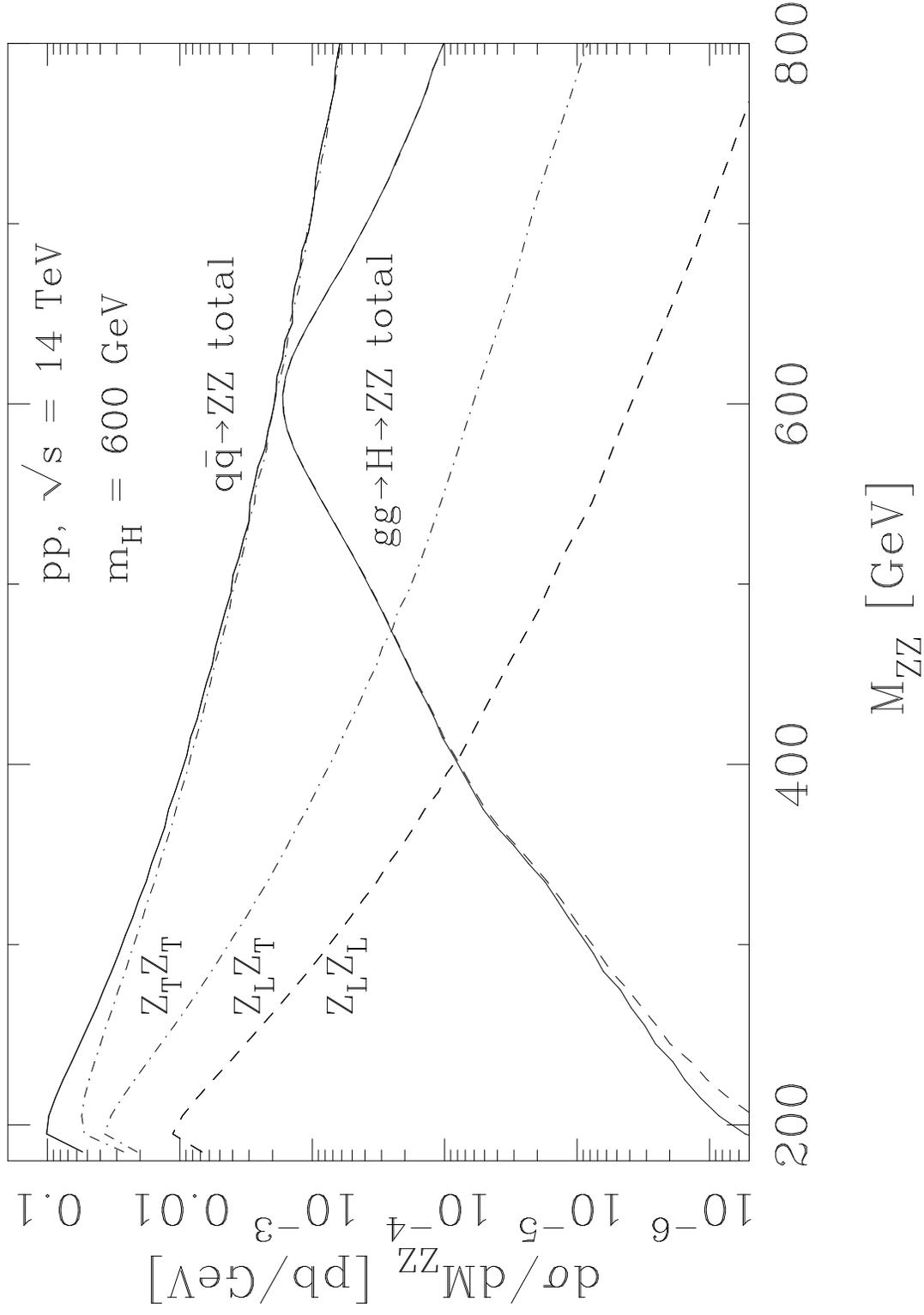
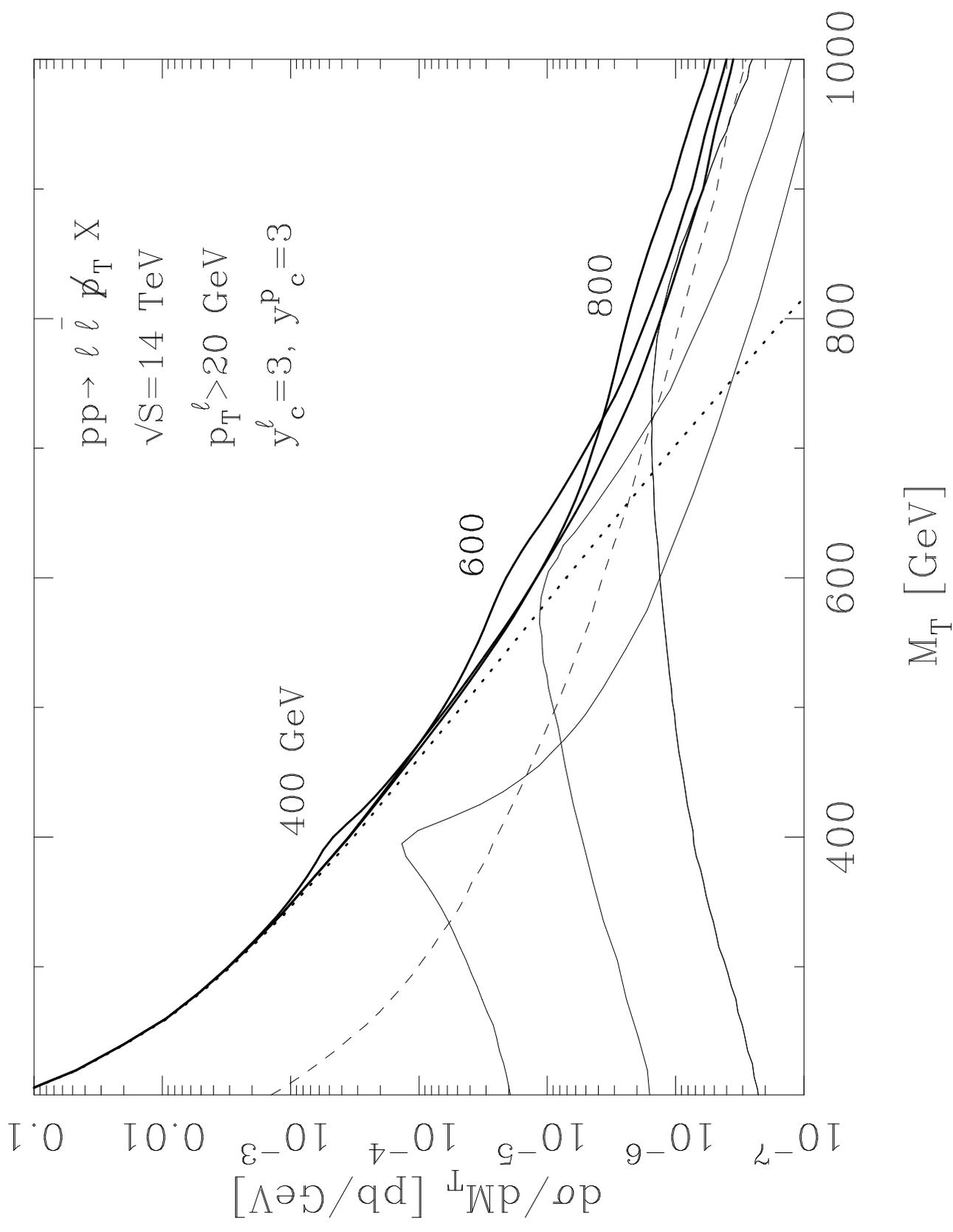
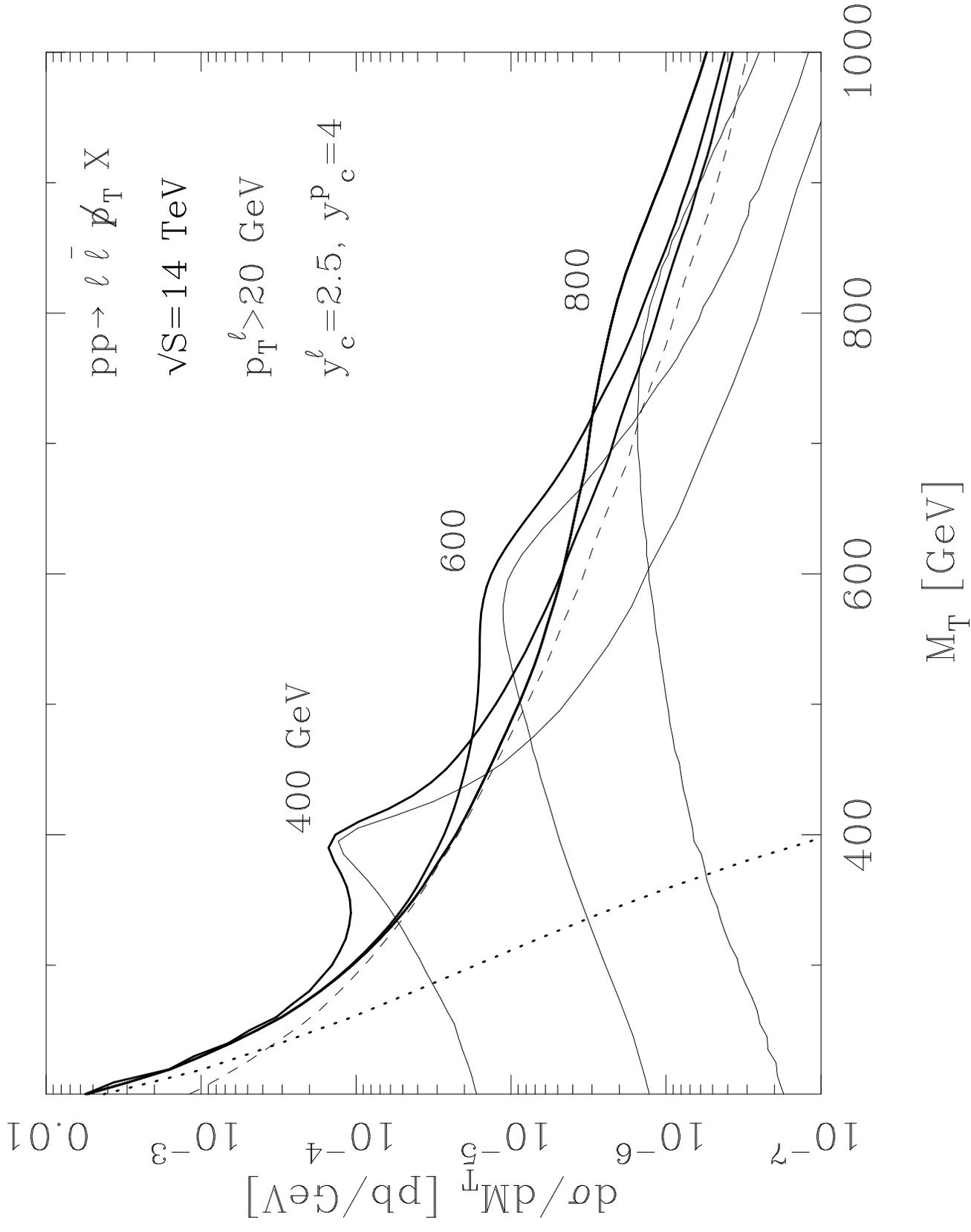
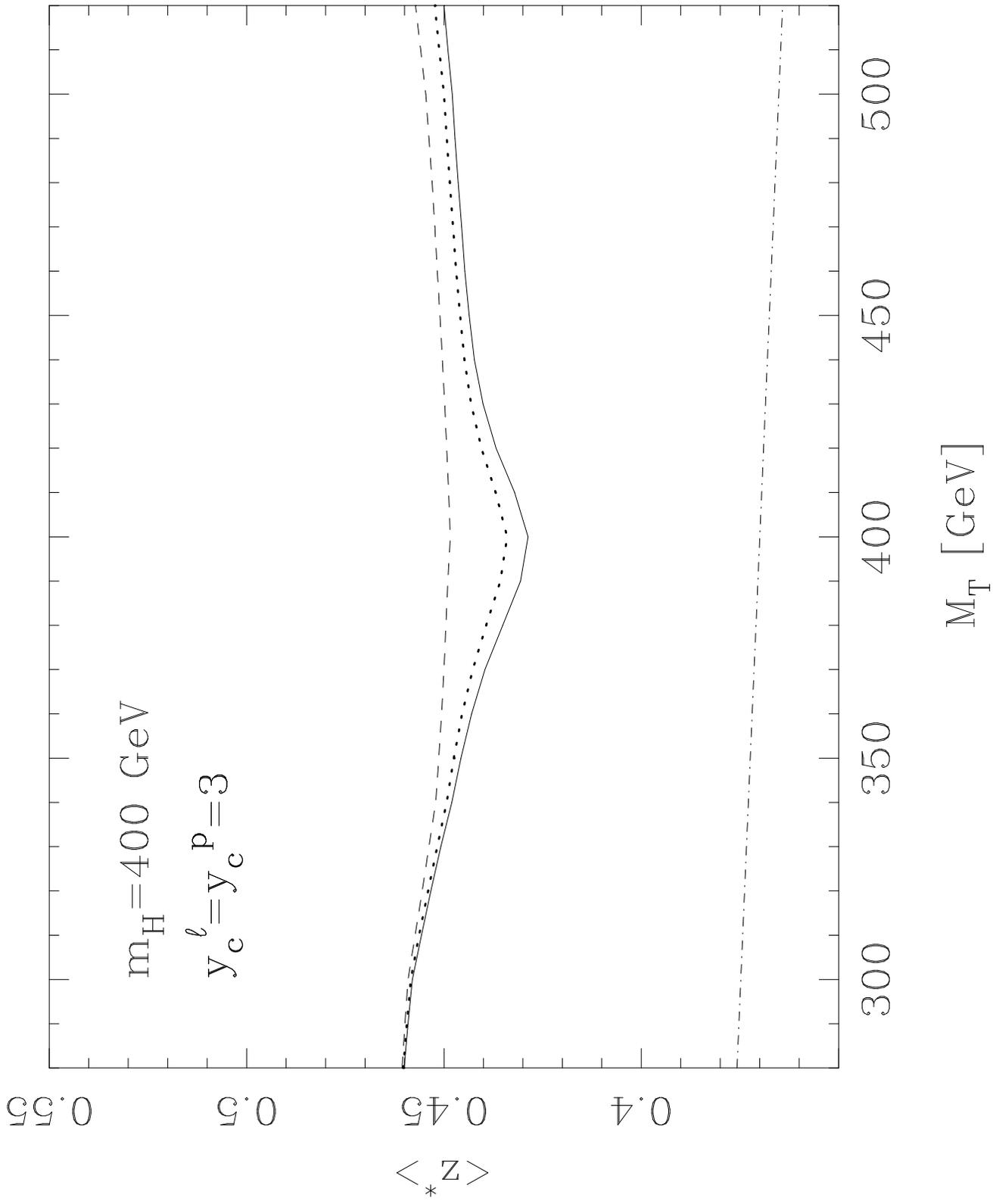
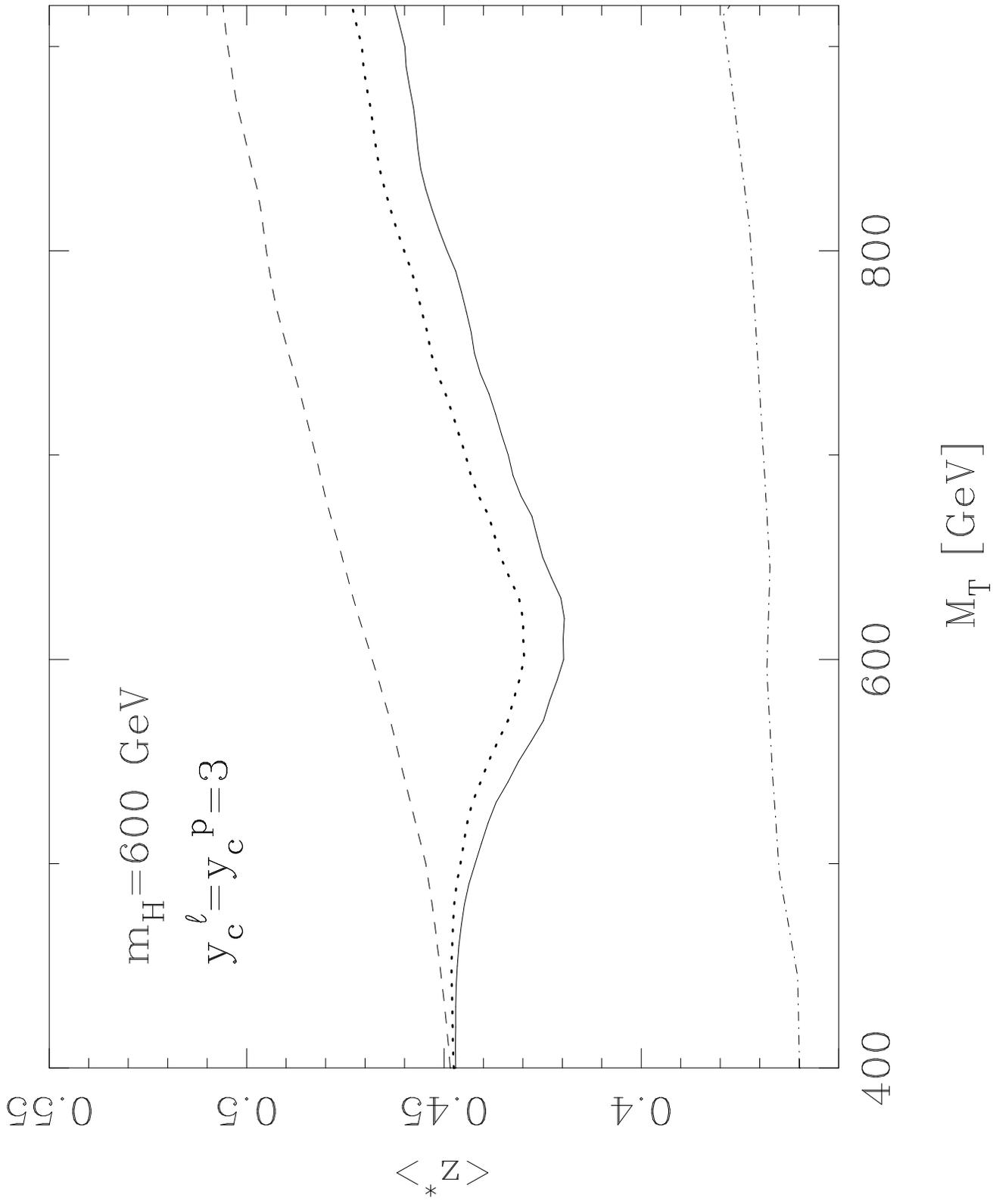
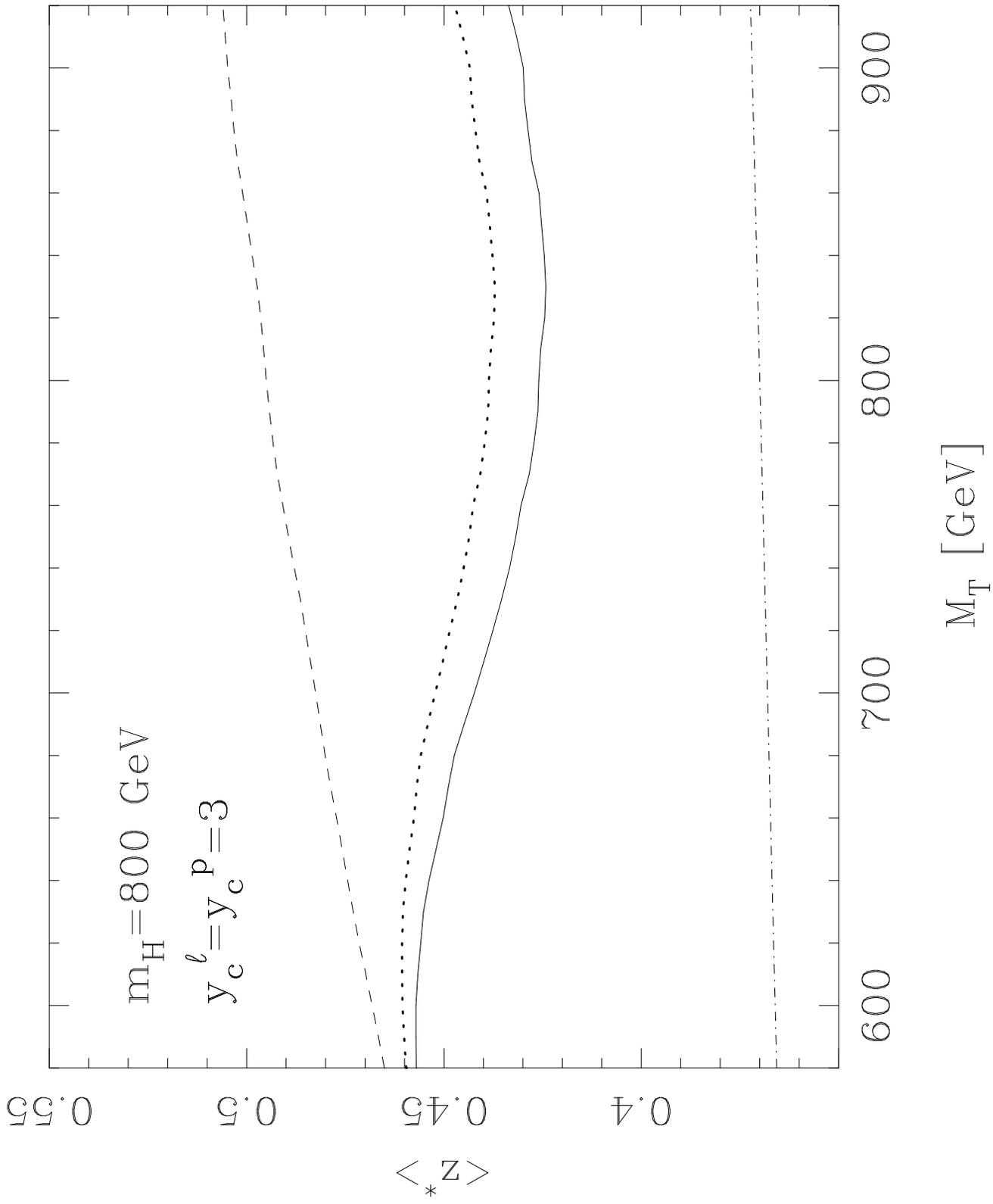
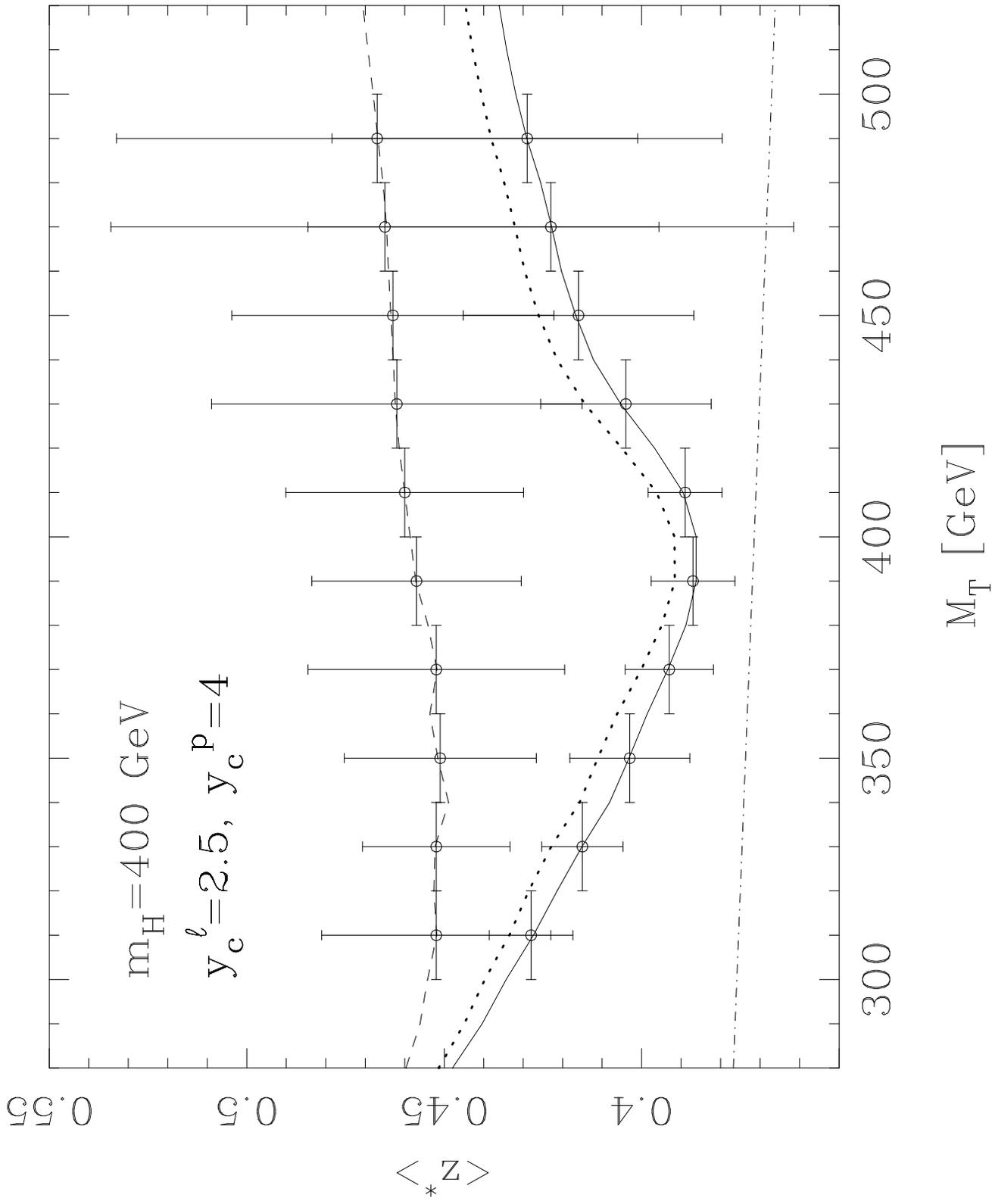
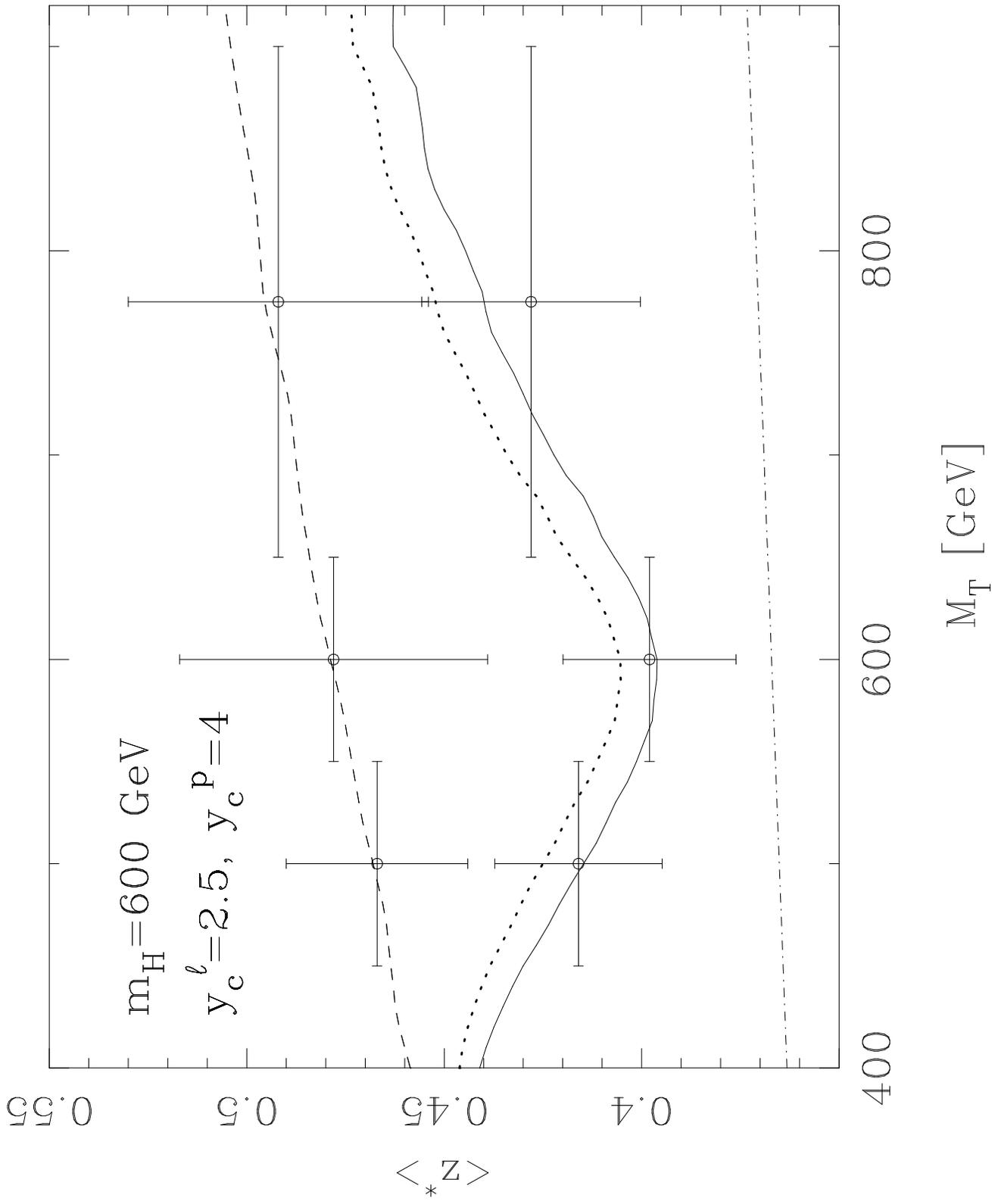
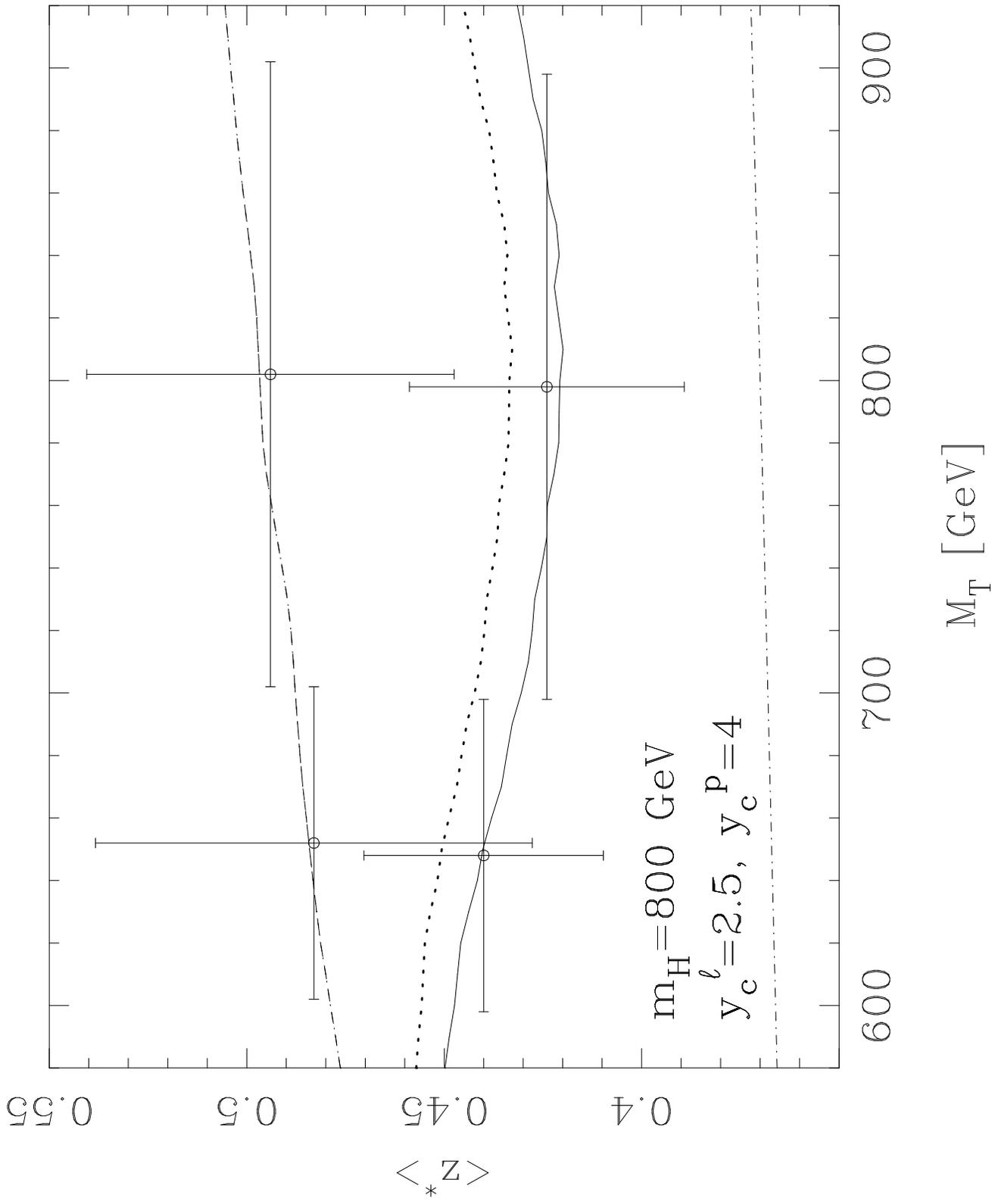
\end{document}